\documentclass[12pt,preprint]{aastex}

\newcommand{\sat}[1]{\it\uppercase{#1}\rm}
\newcommand{\fig}[1]{Fig.~\ref{#1}}

\newcommand{\aspeed}[1]{$\sim$#1 km s${}^{-1}$}
\newcommand{\rsun}[1]{${#1}\,R_\odot$}

\begin{document}

\shorttitle{Current Sheet} %

\shortauthors{Liu et al.}

\title{A Reconnecting Current Sheet Imaged in A Solar Flare}

\author{\small Rui Liu\altaffilmark{1},Jeongwoo Lee\altaffilmark{2}, Tongjiang
Wang\altaffilmark{3}, Guillermo Stenborg\altaffilmark{4},  Chang Liu\altaffilmark{1}, \& Haimin
Wang\altaffilmark{1} \normalsize}

\altaffiltext{1}{Space Weather Research Laboratory, Center for Solar-Terrestrial Research, NJIT,
Newark, NJ 07102; rui.liu@njit.edu}

\altaffiltext{2}{Department of Physics, NJIT, Newark, NJ 07102}

\altaffiltext{3}{Catholic University of America and NASA Goddard Space Flight Center, Greenbelt, MD
20771}

\altaffiltext{4}{Interferometrics, Inc. Herndon, VA 20171}

\begin{abstract}

Magnetic reconnection changes the magnetic field topology and powers explosive events in
astrophysical, space and laboratory plasmas. For flares and coronal mass ejections (CMEs) in the
solar atmosphere, the standard model predicts the presence of a reconnecting current sheet, which
has been the subject of considerable theoretical and numerical modeling over the last fifty years,
yet direct, unambiguous observational verification has been absent. In this Letter we show a bright
sheet structure of global length ($>0.25\ R_\sun$) and macroscopic width ((5--10)${}\times10^3$ km)
distinctly above the cusp-shaped flaring loop, imaged during the flare rising phase in EUV. The
sheet formed due to the stretch of a transequatorial loop system, and was accompanied by various
reconnection signatures that have been dispersed in the literature. This unique event provides a
comprehensive view of the reconnection geometry and dynamics in the solar corona.

\end{abstract}


\section{Introduction}

A vertical current sheet is expected to form above the flare loop when a closed magnetic structure
is highly stretched in solar conditions \citep{kp76, karpen95, lf00, linker03}. Pieces of indirect
evidence highly suggestive of such a reconnection geometry have accumulated over decades of
observations \citep{pf02}. These include candlelight flare loops implying an X-type or Y-type
reconnection point above the cusp \citep{tsuneta92}, high-temperature plasma along the field lines
mapping to the tip of the cusp \citep{tsuneta96a}, loop shrinkage implying the relaxation of newly
reconnected field lines \citep{fa96}, a hard X-ray (HXR) source above the soft X-ray loop top
\citep{masuda94}, upward-moving plasmoid \citep{shibata95} and supra-arcade downflows \citep{mh99}
implying reconnection outflows, horizontal converging motion above a cusp-shaped flare loop
implying reconnection inflows \citep{yokoyama01}, and a double coronal source morphology in HXRs
implying the formation of a current sheet between the two sources \citep{sh03}. Recently,
considerable attention has been given to a coaxial, bright ray feature that appears several hours
after some CMEs \citep{ciaravella02, ko03, raymond03, webb03, lin05, bemporad06, cr08, lin09}. The
flaring-time current sheet, however, has not been directly detected yet.

In this Letter, we study a transequatorial loop system (TLS), whose eruption on 2004 July 29
resulted in a halo CME and a \sat{goes}-class C2.1 flare. Most importantly, a bright elongated
feature is observed to extend above a cusp-shaped flare loop during the impulsive phase, whose
geometry and dynamics are highly suggestive of a reconnecting current sheet.

\section{Observation}

\begin{figure} \epsscale{0.80}
\plotone{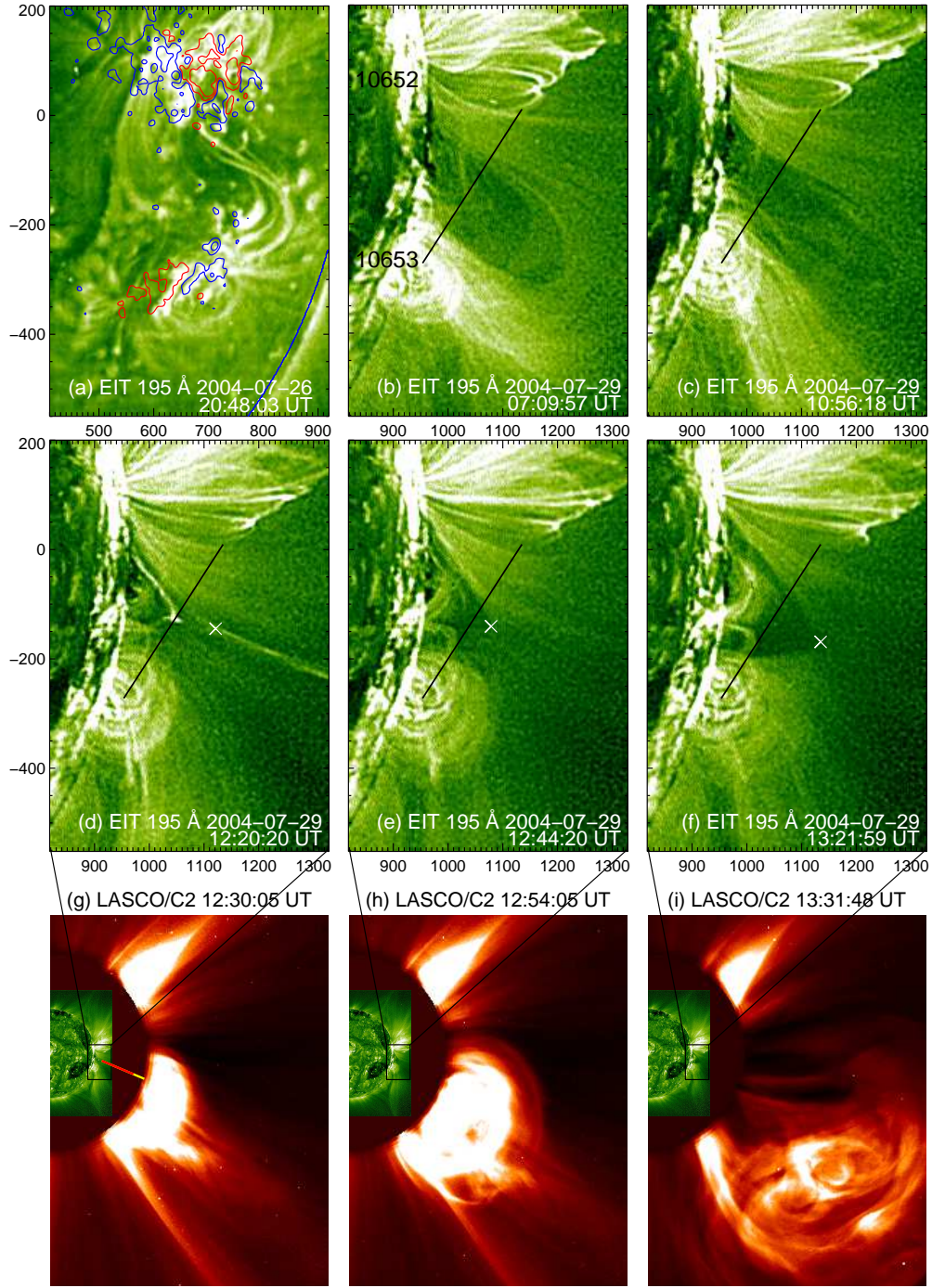} \caption{Rising and subsequent eruption of the TLS in EUV and white-light.
West is to the right and north is to the top. The unit in X- and Y-axes is arcsec in Panels
(a)--(f). In Panel (a) the EIT image is overlaid with a magnetogram taken by the \sat{soho}
Michelson Doppler Imager at approximately the same time. Contours levels are $\pm100$ and $\pm500$
G, with red and blue colors indicate positive and negative polarities, respectively. The Y-point is
marked by `x' in Panels (d)--(f). In Panel (g) the lower and upper bounds of the current-sheet
length inferred from radio emissions are indicated by the line in red and in yellow, respectively.
\label{eit}}
\end{figure}

\subsection{Instruments}

The TLS connected NOAA AR 10652 to AR 10653 (\fig{eit}(a) and (b)). During its disk passage, the
TLS was oriented primarily in the north-south direction, with the northern footpoints located to
the east of the southern ones by $\sim10^\circ$. When rotated with the Sun to the west limb, the
TLS erupted as a halo CME associated with a flare at about 12:00 UT on 2004 July 29, observed by
the EUV Imaging Telescope \citep[EIT;][]{delab95} on board the Solar and Heliospheric Observatory
(SOHO). EIT takes full-disk images at a pixel scale of $2.6''$ pixel${}^{-1}$ and 12-min cadence in
a narrow bandpass centered on 195 {\AA} (Fe~XII; $1.6\times10^6$ K). The 195 {\AA} channel is also
sensitive to high-temperature flare plasma \citep{tripathi06}, because of the presence of the
Fe~XXIV resonance line ($\lambda$192; $2\times10^7$ K) . In addition to the standard EIT data
processing, we have further removed the instrumental stray-light background, and enhanced the fine
coronal structures in EIT images with a wavelet method \citep{svh08}. The Coronal Diagnostic
Spectrometer \citep[CDS;][]{harrison95} on board \sat{soho} was also pointing at AR 10652. A raster
scan from 13:32 to 15:32 UT covered the northern leg of the post-flare loop, despite the limited
field of view (FOV; $4'\times4'$) of the raster images. The ensuing CME was observed in white-light
by the Large Angle and Spectrometric Coronagraph \citep[LASCO;][]{brueckner95} on board SOHO, which
consists of two optical systems, C2 (2.2--6.0 $R_\sun$) and C3 (4--32 $R_\sun$). Relevant radio
emissions were recorded by the Green Bank Solar Radio Burst Spectrometer (GBSRBS) on the ground and
the WAVES instrument on board the WIND spacecraft \citep{bougeret95}.

\subsection{Analysis}

\begin{figure}\epsscale{0.7}
\plotone{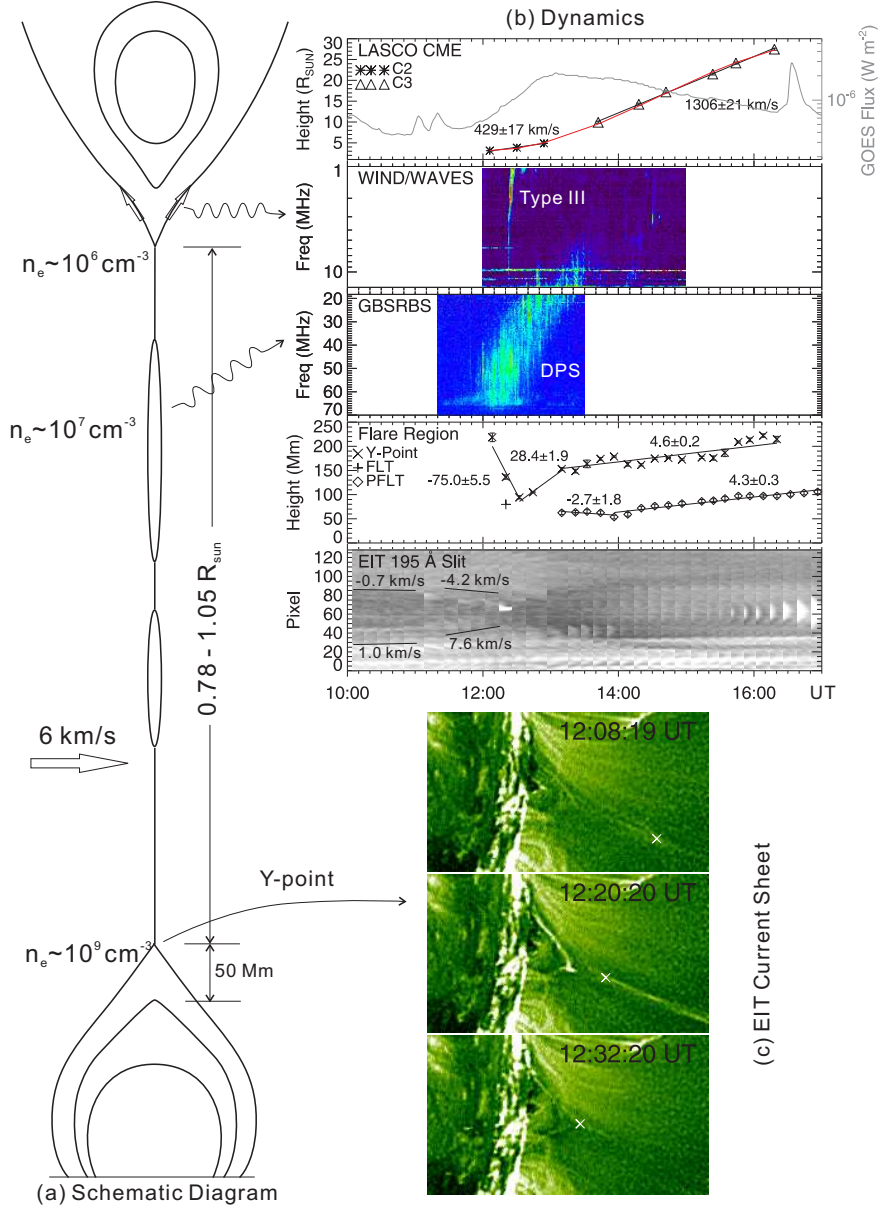} \caption{Geometry and dynamics of the reconnection region. (a) A schematic
diagram illustrating the reconnection geometry. Plasma densities at the upper tip of the current
sheet ($\sim\,$10${}^6$ cm${}^{-3}$) as well as in the plasmoid ($\sim\,$10${}^7$ cm${}^{-3}$) are
derived from the radio emissions. The density of the flare loop ($\sim\,$10${}^9$ cm${}^{-3}$) is
inferred from density-sensitive line pairs observed by CDS. (b) From top to bottom: height-time
profile of the LASCO CME (asterisks), and the GOES 1--8 {\AA} SXR flux (grey color); WIND WAVES
spectrogram (1.08--13.83 MHz) showing a Type III burst; GBSRBS spectrogram (18.29--69.96 MHz)
showing drifting pulsating structures; height-time evolution of the Y-point (`x'), the flare
looptop (FLT; `+'), and the post-flare looptop (PFLT; diamond) as well as their average speed in km
s${}^{-1}$; chronological observation of the transequatorial loop through the slit in
\fig{eit}(a--f). (c) Initial downward motion of the Y-point, marked by an `x' symbol; the field of
view is $585''\times350''$. \label{cartoon}}
\end{figure}

The TLS that erupted on 2004 July 29 12:00 UT can be clearly seen to slowly rise from as early as
2004 July 28 21:12 UT. With footpoints being ``fixed'' on the extremely dense photosphere, the
``waist'' of the TLS became thinner and thinner (\fig{eit}(b)--(c)). By placing a slit across the
waist (\fig{eit}(b)--(f)), and then putting the resultant strips in chronological order (bottom
panel of \fig{cartoon}(b)), one can see that initially the waist converged at \aspeed{1}, and then
the speed suddenly increased to $4.2\sim7.6$ km s${}^{-1}$ from 11:08 until 12:20 UT when a
cusp-shaped flare loop formed \citep[\fig{eit}(d);][]{tsuneta96a}. Meanwhile, from 11:30 till 13:30
UT, GBSRBS recorded drifting pulsating structures (DPSs) at metric frequencies (3rd panel of
\fig{cartoon}(b)), suggestive of the tearing of a current sheet \citep[\fig{cartoon}(a);][]{kkb00}.
In this scenario, electrons are accelerated and trapped as the plasmoids (magnetic islands)
contract and mutually interact, probably in a classic Fermi manner \citep{drake06}, which generate
the individual pulses of the DPSs; motions of the plasmoids in the corona result in the global
drift of the DPSs. With reconnection sets in, we expect that the plasma near the reconnection
region would be rapidly heated to flaring temperatures in excess of $10^7$ K, which would make the
reconnection region a bright feature in EUV lines of highly ionized lines, such as Fe XXIV covered
by the EIT 195 {\AA} filter.

Indeed, at 12:20 UT (\fig{eit}(d)), a bright, collimated feature can be seen to extend from well
above a cusp-shaped flare loop for 170 Mm (\rsun{0.25}) up to the border of the EIT FOV
(\rsun{1.5}), indicating that the TLS had been stretched to the point that the oppositely directed
field lines at the waist were close enough to reconnect. The morphology is similar to that in the
standard model \citep{kp76}, and the extended sheet structure favors a Y-type \citep{lf00} over an
X-type geometry \citep{ys01}. The sheet spans about 3--5 EIT pixels, i.e., (5--10)$\times10^3$ km.
Hence the observed thickness is not resolution-limited since the EIT point spread functions are
narrower than the pixel size of the CCD \citep{delab95}. However, projection effects could increase
the apparent thickness by a factor of 2--4 \citep{lin09}. The (inverse) Y-point, i.e., the lower
tip of the sheet, is above the cusp-shaped flare looptop by about 50 Mm, with the height ratio of
about 1.6 between the Y-point and the looptop.  From 12:08 to 12:32 UT, the lower tip of the sheet
(\fig{cartoon}(b--c)) moved toward the solar surface. Meanwhile ($\sim\,$12:23 UT), the WIND
spacecraft recorded a Type III burst (2nd panel of \fig{cartoon}(b)), indicative of accelerated
electrons beaming upward from the upper tip of the current sheet \citep[\fig{cartoon}(a);][]{ml85}.
One can see from the radio spectrograms that the starting frequency of the Type III burst, $13.8\
\mathrm{MHz}\leq f<18.3\,\mathrm{MHz}$. By exploiting the density profile known as the
Baumbach-Allen formula \citep{cox01}, one obtains that $1.95\,R_\sun < R \leq 2.22\,R_\sun$, where
$R$ is the heliocentric distance of the upper tip of the current sheet. The heliocentric distance
of the lower tip, $r$, is measured to be \rsun{1.19}; and the sheet orientation is deviant from the
radial direction by $15^\circ$, hence the sheet length, $L$, can be estimated, i.e., $0.78\,R_\sun
< L \leq 1.05\,R_\sun$ (\fig{eit}(g)).

\begin{figure}\epsscale{0.8}
\plotone{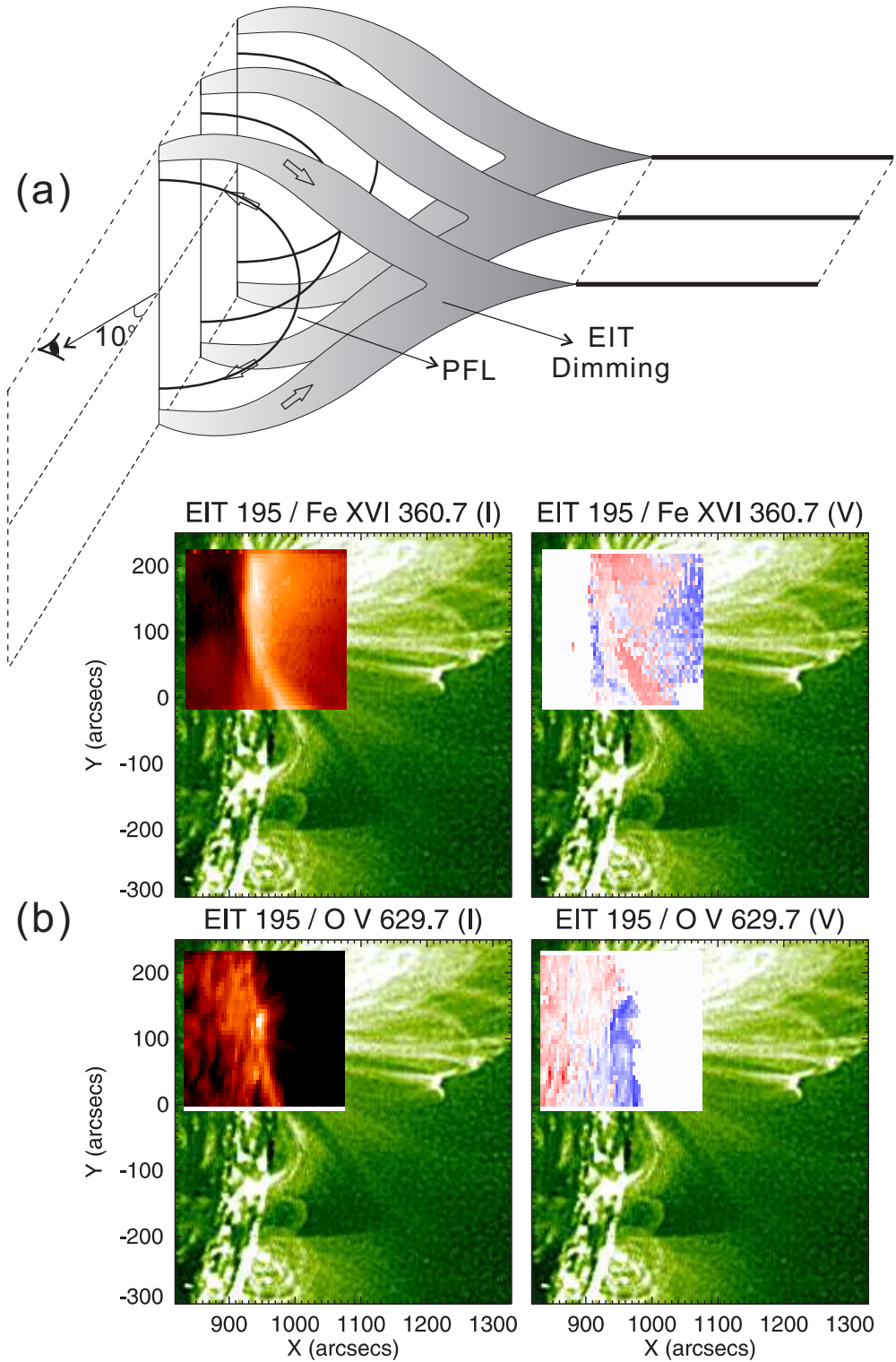} \caption{Upflow and downflow detected by CDS. (a) A schematic diagram
illustrates the face-on view of the reconnection region; (b) CDS intensity images of Fe XVI and O V
(left column) and the corresponding velocity maps (right column) overlaid on the same EIT 195 {\AA}
image taken at 2004 July 29 13:32 UT.\label{doppler}}
\end{figure}

The sheet and the cusp-shaped flare loop are visible probably due to the Fe~XXIV emission
($\lambda192$; $2\times10^7$ K). Both faded out quickly after the temporary appearance, presumably
due to cooling. The flare loop reappeared at 13:09 UT at a lower altitude by about 17 Mm, taking a
relaxed shape (e.g., \fig{eit}(f)), above which appeared an emission-depressed, cusp-shaped region
(\fig{eit}(e)--(f)). Note that the bright loops apparently within the southern leg of the dimming
region are not newly formed but relatively long-lived active-region loops of AR 10653, which may be
located in the foreground or background due to the optically thin nature of coronal lines. The EIT
cusp-shaped dimming is in emission in the hot Fe~XVI line ($\lambda360.8$; $2.7\times10^6$ K; top
left panel of \fig{doppler}(b)), while remaining dark in the cold O~V line ($\lambda629.7$;
$2.5\times10^5$ K; bottom left panel of \fig{doppler}(b)). Clearly, the cusp-shaped region hosts
newly reconnected flux tubes that later cooled and relaxed into the dipolar post-flare loops
\citep{fa96}. Velocity maps (right column of \fig{doppler}(b)) show that Doppler redshifts dominate
in the cusp-shaped region in Fe~XVI, while blueshifts dominate in the post-flare arcade in O~V.
This is because the flare-loop plane is tilted eastward by about $10^\circ$ as illustrated in
\fig{doppler}(a), hence upflows due to evaporated chromospheric plasma filling the newly
reconnected flux tubes bear a velocity component away from the observer (redshift) at the northern
leg of the cusp-shaped region; meanwhile, downflows due to cooled plasma falling back along the
relaxed field bear a velocity component toward the observer (blueshift) at the northern leg of the
post-flare arcade. As an important consequence of the reconnection in the corona, the evaporation
takes place when chorompospheic material is heated by precipitating nonthermal electrons or thermal
conduction \citep{pf02}. Both the cusp-shaped dimming and the underlying post-flare arcade expanded
with time (\fig{eit}(e)--(f)), implying reconnection proceeding to higher and higher altitudes
\citep{pf02}.

\begin{figure}\epsscale{0.9}
\plotone{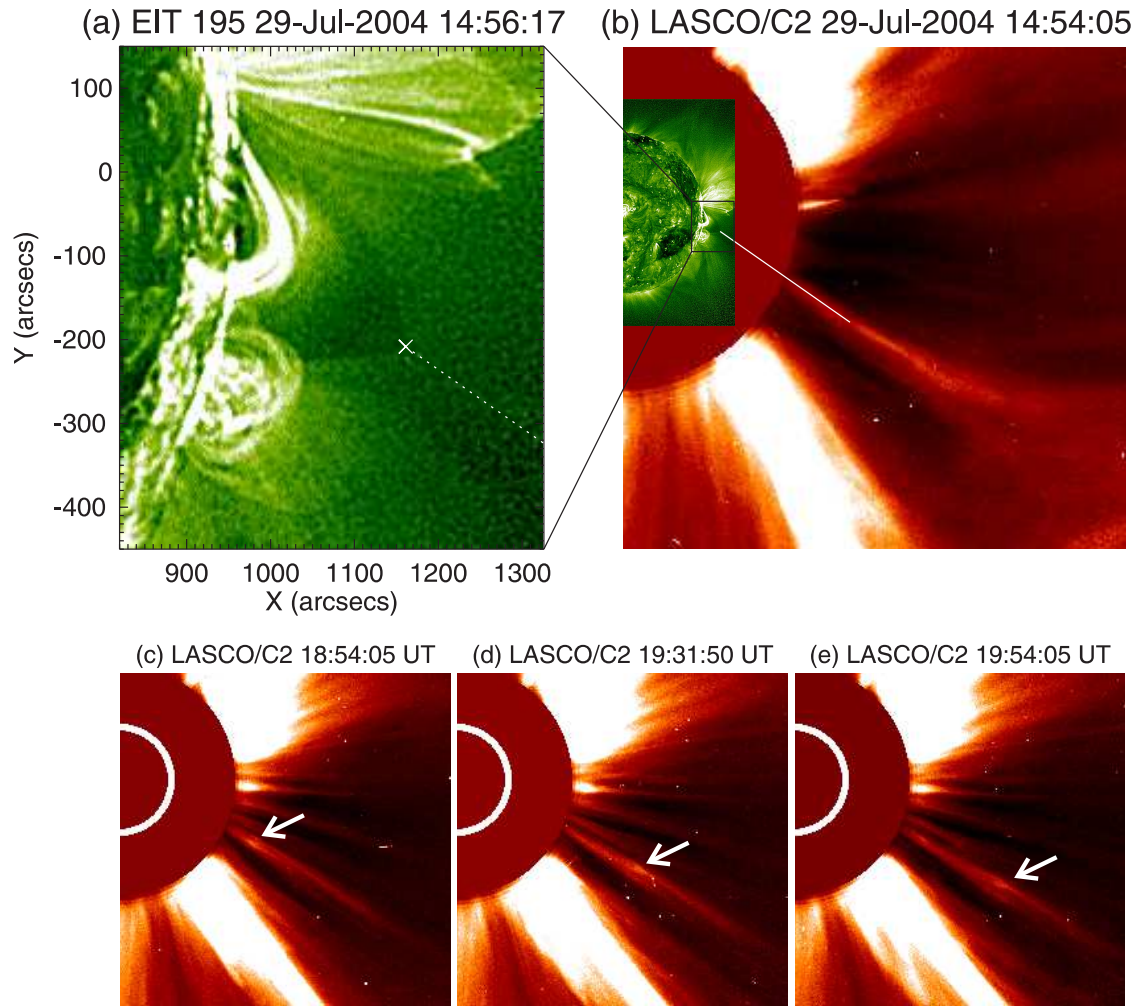} \caption{Post-CME ray observed in LASCO C2 (a) in relation to the cusp-shaped
dimming in EIT 195 {\AA} (b). (c--e) A blob moving outward along the post-CME ray (marked by
arrows), at a speed of \aspeed{360}.\label{ray}}
\end{figure}

Despite the low contrast, one can still get a sense of the orientation of the current sheet from
the cusp morphology. It was apparently aligned with a post-CME ray feature observed in coronagraph
(top panels of \fig{ray}). The ray was seen to rapidly fan out above its upper tip (\fig{ray}(b))
in a similar fashion as the vertical current sheet does in the models \citep{kp76, lf00}, which is
not reported in previous observations \citep[e.g.,][]{webb03, ko03, lin05}. A blob was observed to
move outward along the ray at \aspeed{360} (bottom panels of \fig{ray}), implying the ejection of a
plasmoid.

These observations allow us to estimate the rate of magnetic reconnection, $M_R$, given by the
ratio between the inflow and the outflow speed \citep{pf00}. If we relate the ejected blob to the
reconnection outflow \citep{ko03,lin05}, and assume that the reconnection inflow can be
characterized by the converging of the loop waist \citep[\fig{eit} and the bottom panel of
\fig{cartoon}(b);][]{yokoyama01}, then $M_R=(4.2\sim7.6)/360\approx{}$0.01--0.02, despite the
caveat that the outflow is not simultaneous to the inflow. According to the steady reconnection
theory \cite{pf00}, the sheet thickness $d=L\times M_R$ for a diffusion region of length $L$ and
width $d$. With $0.78\,R_\sun < L \lesssim 1.05\,R_\sun$ and $M_R={}$0.01--0.02, we find that
$d\approx{}$(5--15)${}\times10^3$ km, in good agreement with the observed sheet thickness
((5--10)${}\times10^3$ km). The dissipation of the current sheet may help further accelerate the
CME to a speed (\aspeed{1300}; top panel of \fig{cartoon}(b)) much faster than the one
(\aspeed{400}) when the current sheet was first detected. A follow-up paper is being prepared on
more detailed properties of this reconnection region.

\section{Conclusion \& Discussion}

The various reconnection signatures observed in this single event, including plasma inflow and
outflow, an expanding cusp-shaped dimming, chromospheric evaporation, and the relevant radio
emissions, constitute for the first time a comprehensive view of the reconnection dynamics in the
solar corona. Taking its formation, geometry and dynamics into account, we conclude that the
bright, elongated EUV feature observed above the cusp-shaped flaring loop is a Y-type current
sheet.

This unique observation has several important implications. First of all, the Y-type reconnection
geometry together with the above dynamic features constitutes direct verification of the standard
physical picture for flares/CMEs \citep{kp76,lf00}, which has been invoked in other astrophysical
phenomena beyond stellar flares, e.g., episodic jets from black hole systems \citep{yuan09}.
Second, the fact that the lower tip of the current sheet is located well above the cusp-shaped
flare loop provides an unequivocal evidence that particle acceleration in solar flares occurs well
above the flare loop. The reconnection geometry observed here agrees with an independent study of
electron time-of-flight \citep[TOF;][]{aschwanden96}, in which a scaling law is found between the
electron TOF distance $l'$, indicating the height of the acceleration site, and the flare-loop half
length $s$, i.e., $l'/s=1.4\pm0.3$. Third, the fact that the lower tip of the current sheet
initially moved downward supports the conjecture that the descending motion of the flare looptop
emission during the flare rising phase, which the standard flare model fails to explain, is due to
the extending of a current sheet \citep{sh03}. In our case, the plasma inflows due to the stretch
of the TLS may bring in flux faster than the rate of dissipation. With the flux piling up, the
diffusion region grows in length in both (upward and downward) directions until it reaches the
global dimension ($R_\sun$) and becomes subject to tearing \citep{biskamp86}. Fourth, the imaging
observation allows for a direct measurement of the thickness of the flaring-time current sheet. The
measured width ((5--10)${}\times10^3$ km) is an order of magnitude thinner than that of post-CME
rays \citep[$10^5$ km;][]{ciaravella02, ko03, cr08, lin09}. However, it agrees with the steady
reconnection theory \citep{pf00}, and matches not only the width ($\sim\,$4${}\times10^3$ km) of
high-speed outflows of hot plasma near a reconnection site, detected in EUV spectra \citep{wsq07},
but also the thickness ((3--10)${}\times10^3$ km) of the heliospheric current sheet measured
\emph{in situ} \citep{winterhalter94}. Thus, we have obtained a tight upper limit to the ``true''
thickness of the current sheet in the solar corona. Finally, the observation of a current sheet of
global length and macroscopic width, which is associated with radio pulsations \citep{kkb00},
supports recent theoretical studies which conclude that a single localized reconnection region
cannot account for the large number of energetic electrons typically seen in flares
\citep[e.g.,][]{egedal09}.

\acknowledgments SOHO is a project of international cooperation between ESA and NASA. R.~L., C.~L.,
and H.~W. was supported by NASA grant NNX08-AJ23G and NNX08-AQ90G, and by NSF grant ATM-0849453.
J.~L. was supported by NSF grant AST-0908344. T.~W. was supported by NASA grant NNX08AP88G and
NNX09AG10G.


\end{document}